# On dynamics of imploding shock waves in a mixture of gas and dust particles

R. K. Anand

**Department of Physics, University of Allahabad, Allahabad-211002, India**

*E-mail address:* **anand.rajkumar@rediffmail.com**

**Abstract**

In this paper, the generalized analytical solutions for one-dimensional adiabatic flow behind the imploding shock waves propagating in a dusty gas are obtained using the geometrical shock dynamics theory. The dusty gas is assumed to be a mixture of a perfect gas and spherically small solid particles, in which solid particles are continuously distributed. Anand's shock jump relations for a dusty gas are taken into consideration to explore the effects due to an increase in (i) the propagation distance from the centre of convergence, (ii) the mass fraction of solid particles in the mixture and (iii) the ratio of the density of solid particles to the initial density of the gas, on the shock velocity, pressure, temperature, density, velocity of mixture, speed of sound, adiabatic compressibility of mixture and the change-in-entropy across the shock front. The results provided a clear picture of whether and how the presence of solid particles influences the flow field behind the imploding shock front.

## 1. Introduction

The gas-particle two-phase flows occur in a variety of natural phenomena and are involved in many industrial processes. The natural phenomena accompanied by the gas-particle two-phase flows are typified by explosion of supernova, sand storms, moving sand dunes, aerodynamic ablation, cosmic dusts, etc. The flow field, that develops when a moving shock wave hits a two-phase gas-particle medium, has a close practical relation to industrial applications, e.g. solid rocket engine in which aluminium particles are used to reduce the vibration due to instability, as well as industrial accidents such as explosions in coalmines and grain elevators. In heat transfer applications, the gas-particle flows are involved in nuclear reactor cooling and solar energy transport using graphite suspension flows.

So far, a number of papers have been reported on the shock waves in a mixture of gas and dust particles. The structure of gas-particle two-phase flows behind the normal shock waves under different conditions was considered in a large number of papers, starting from the pioneering works of Carrier [1], Rudinger [2] and Kribel [3]. Experimental study of the shock structure in a gas-particle mixture was fulfilled by Outa et al. [4]. The main stages of the investigations of two-phase flows with shock waves are reviewed by Marble [5], Kraiko et al. [6], Higashino and Suzuki [7], Ivandaev et al. [8], Igra and Ben-Dor [9], Pai and Luo [10] and Ben-Dor [11]. The work of Pai [12], Rudinger [13], Pai et al. [14], Miura and Glass [15], Steiner and Hirschler [16], Saito et al. [17], Naidu et al. [18] and Anand [19-20] is worth mentioning in the context of this paper.

One of the main reasons for continuing interest in shock focusing is its ability to create extremely high temperature and pressure at the centre of convergence. However as the strength of imploding shock waves increases the effects due to dust particles become significant. These effects need to be accounted for in order to correctly describe the post shock conditions and acquire information on the attainable pressure and temperature by shock focusing. The study of propagation of shock waves in a dusty gas is of immense significance due to its wide applications to supersonic flights in a dusty gas environment, supersonic-vehicle motion in desert sand storms or clouds of volcanic dust, exhaust plumes of rocket motors propelled by solid fuels, ensuring the explosion safety of coal mines and industrial manufacture of powder materials, coating technologies using supersonic two-phase jets, etc. This has developed our interest in studying the imploding shock waves propagating in a mixture of gas and dust particles. The purpose of writing this paper is, therefore, to present the analytical solutions for one-dimensional adiabatic flow behind the imploding shock waves in a mixture of gas and solid particles to pursue the applications of shock waves in a dusty gas environment. To our best knowledge, so far there is no paper reporting the analytical solutions for imploding shock waves in a dusty gas, obtained by using the geometrical shock dynamics [21-22]. The geometrical shock dynamics approach gives highly accurate results especially, in the case of a spherical symmetry. Since an imploding shock is strengthened as it focuses on the origin.

In the present research paper, the influence of dust particles has been investigated on the flow quantities of the region just behind the imploding shock front. For this purpose, a model based on geometrical shock dynamics is developed to provide a simplified and complete treatment for the propagation of imploding shock waves in a dusty gas during the convergent process. Geometrical shock dynamics was introduced by Whitham [21] and it is important to mention that the original geometrical shock dynamics method does not account for the influence of the flow ahead of the shock front. To obtain the analytical solutions for one dimensional adiabatic flow behind the imploding shock waves in a dusty gas we make the following assumptions and approximations: (i) the dusty gas is a mixture of a perfect gas and small solid particles, (ii) the dust phase comprises the total amount of solid particles which are continuously distributed in the perfect gas, (iii) the gas follows the equation of state of an ideal gas, (iv) the motion of gas can be regarded as inviscid, so that the fluid viscosity and conductivity are neglected except in the interaction with the particles, (v) the solid particles are inert, rigid and spherical, (vi) the particles have constant heat capacity and uniform temperature distribution, (vii) the particle collision is the dominant mechanism of inter-particle interaction. At the modest particle volume fraction to be considered here, particle compaction can be ignored, (viii) the effect of thermal radiation is negligible. It is worth mentioning, however, that a cloud of particles is usually a better emitter and absorber of radiation than a pure gas. Thus, the hot particles downstream of the shock may preheat the cold particles upstream of the shock by radiation. This effect becomes significant as the surface area of the particles increases, (ix) the inter-particle heat transfer due to particle collision is neglected, (x) the gas-particle flow is one-dimensional, (xi) the medium is initially uniform and at rest, and (xii) the disturbances due to the reflections, wave interactions in the wake, etc. do not overtake the imploding shock wave. It is noted that with an increase in the particle volume fraction, particle-particle collisions and other types of particle interactions become increasingly important and thus the assumption of negligible particle interactions is no longer feasible.

The analytical expression for the propagation velocity of shock is obtained by substituting the generalized shock jump relations derived by Anand [20] into the negative characteristic equation. The general non-dimensional forms of analytical expressions for the distribution of pressure, temperature, density, velocity of mixture, speed of sound and adiabatic compressibility of mixture just behind the imploding shock front are obtained, assuming the medium to be initially uniform and at rest. Most of the prior studies have remained focused on the propagation of shock waves in an ideal or dusty gaseous media without discussing the change-in-entropy across the shock front. The expression for change-in-entropy across the imploding shock front is also derived. Two cases are considered–cylindrical and spherical imploding shocks to highlight the differences between the 2D and 3D convergence. The planar case is not of interest since no area convergence and shock amplification exist and it is simply the ordinary planar blast wave problem. The numerical estimations of flow variables behind the imploding front with cylindrical and spherical shock symmetries are carried out using MATHEMATICA and MATLAB codes. The effects of the mass fraction (concentration) of solid particles in the mixture and the dust loading parameter, i.e. the ratio of the density of the solid particles to the initial density of the gas are explored as the imploding shock wave propagates towards the centre of convergence. This model appropriately makes obvious the effects due to an increase in (i) the propagation distance from the centre of convergence, (ii) the mass concentration of solid particles in the mixture and (iii) the ratio of the density of the solid particles to the initial density of the gas, on the propagation velocity of shock, pressure, temperature, density, velocity of mixture, speed of sound, adiabatic compressibility of medium and the change-in-entropy across shock front. The results are displayed graphically and discussed by comparison with those for the case of a perfect or dust-free gas flow. Thus, the results provided a clear picture of whether and how the presence of dust particles influences the flow field behind the imploding shock front.

The paper is organized as follows. The background information is provided in Section 1 as an introduction. Section 2 contains general assumptions and notations as well as shock jump relations. In Section 3 analytical solutions are presented. Section 4 mainly describes results with discussion on the important components of the present model. The last section 5 presents the concluding remarks.

**2. Basic equations and shock jump relations**

The unsteady, one-dimensional flow field in a mixture of a perfect gas and small solid particles is a function of two independent variables; the time $t$ and the space coordinate $r$. The conservation equations for one-dimensional unsteady flow of a dusty gas can be expressed conveniently in Eularian coordinates [20] as follows:

$$\frac{\partial u}{\partial t} + u\frac{\partial u}{\partial r} + \frac{1}{\rho}\frac{\partial p}{\partial r} = 0, \tag{1}$$

$$\frac{\partial \rho}{\partial t} + u\frac{\partial \rho}{\partial r} + \rho\left(\frac{\partial u}{\partial r} + D\frac{u}{r}\right) = 0, \tag{2}$$

$$\frac{\partial e}{\partial t} + u\frac{\partial e}{\partial r} - \frac{p}{\rho^2}\left(\frac{\partial \rho}{\partial t} + u\frac{\partial \rho}{\partial r}\right) = 0, \tag{3}$$

where $u(r,t)$ is the velocity of mixture, $\rho(r,t)$ the density of mixture, $p(r,t)$ the pressure of mixture, $e(r,t)$ the internal energy of mixture per unit mass, $r$ is the distance from the origin O and $t$ is the time coordinate. The dimensionality index $D$ is defined by $D = d\ln A/d\ln r$, where $A(r) = 2\pi D r^D$ is the flow cross-section area. The index $D = 1$, or 2 is for one-dimensional cylindrical, or spherical geometry, respectively.

Dust particles entrained in non-steady gas flow cannot immediately follow the changes in velocity and temperature of the gas. Consequently behind a shock front in a dusty gas there is a 'relaxation zone' in which drag and heat transfer act to bring gas and dust velocities close together again. Due to the condition of velocity and temperature equilibrium, the terms of drag force and heat-transfer rate, which can be expressed via the drag coefficient and the Nusselt number, do not appear in the right-hand sides of the Eqs. (1) and (3). These terms are, of course, important for evaluating the extent of the relaxation zone behind the shock front, which is however, beyond the scope of this paper. It is worth mentioning that the dusty gas is a pure perfect gas which is contaminated by small solid particles and not as a mixture of two perfect gases. The solid particles are continuously distributed in the perfect gas and in their totality are referred to as dust. It is assumed that the dust particles are highly dispersed in the gas phase such that the dusty gas can be considered as a continuous medium where the conservation Eqs. (1) – (3) apply. All relaxation processes are excluded such that no relative motion and no temperature differences between perfect gas and solid particles occur. The solid particles are also assumed to have no thermal motion, and, hence they do not contribute to the pressure of the mixture. As a result, the pressure $p$ and the temperature $T$ of the entire mixture satisfy the thermal equation of state of the perfect gas partition. The equation of state of the mixture subject to the equilibrium condition, is given as

$$p = \left(\frac{1-k_p}{1-Z}\right)\rho R_i T, \qquad (4)$$

where $k_p = m_{sp}/m$, is the mass concentration of solid particles ($m_{sp}$) in the mixture ($m$) taken as a constant in the whole flow field, $Z$ is the volumetric fraction of solid particles in the mixture, $R_i$ is the gas constant and, $T$ is the temperature of the mixture. The relation between $k_p$ and $Z$ is given by Pai et al. [14] as follows:

$$k_p = \frac{Z\rho_{sp}}{\rho}, \qquad (5)$$

where $Z = Z_o \rho/\rho_o$, while $\rho_{sp}$ is the species density of the solid particles and a subscript '*o*' refers to the initial values of $Z$, and $\rho$. It is notable that in equilibrium flow, the mass concentration of solid particles $k_p$ is a constant in the whole flow-field. As a result, $Z/\rho =$ constant, in the whole flow-field. The mass concentration of solid particles $k_p = 0$ refers the case of a perfect gas, i.e. dust-free case. The initial volume fraction $Z_o$ of the solid particles is, in general, not constant. But the volume occupied by the solid particles is very small because the density of the solid particles is much larger than that of the gas [15], hence, $Z_o$ may be assumed as a small constant. The initial volume fraction of the small solid particles is given by Pai [12] as

$$Z_o = \frac{V_{sp}}{V_{go} + V_{sp}} = \frac{k_p}{G(1-k_p) + k_p}, \qquad (6)$$

where the volume of the mixture $V$ is the sum of the volume of the perfect gas at the reference state $V_{go}$ and the volume of the particles $V_{sp}$ which remains constant. The volumetric parameter $G$ is defined as $G = \rho_{sp}/\rho_{go}$, which is equal to the ratio of the density of the solid particles to the initial density of the gas. Hence, the fundamental parameters of the present model are $k_p$ and $G$ which describe the effects of the dust loading. For the dust loading parameter $G$, we have a range of $G=1$ to $G \to \infty$, i.e. $V_{sp} \to 0$. Fig. 1 shows the variation of $Z_o$ the initial volume fraction of small solid particles with $k_p$ the mass concentration of solid particles for the values of $G$ from 1 to $\infty$.

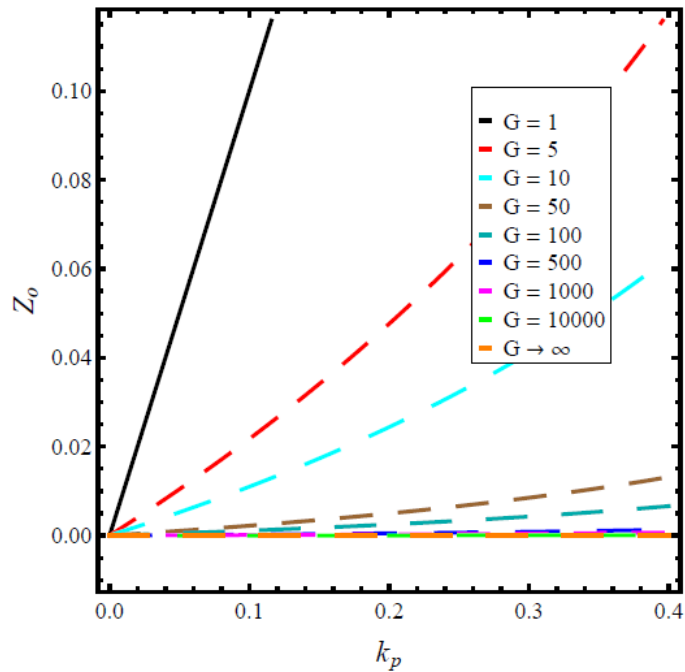

**Fig.1** Variation of $Z_o$ with $k_p$ for various values of $G$.

It is notable that $Z_o$ remains almost unaffected with increase in $k_p$ for large values of $G$, the dust loading parameter. It is worth mentioning that if $G=1$ then $Z_o = k_p$. In this case, the dust and the perfect gas are virtually indiscernible concerning their specific volume, but the volume fraction of the dust still behaves as an incompressible solid phase. The addition of solid particles does not increase the inertia of mixture, and does not additionally slow down any wave propagation. On the other hand, the initial volume fraction of small solid particles $Z_o$ of the dust linearly increases with $k_p$, as the volumetric extension $V_{sp}$ of the dust increases. The dust is considered as an incompressible solid phase. The occurrence of an incompressible phase in the mixture basically lowers the compressibility of the mixture.

The internal energy of the mixture is related to the internal energies of the two species and may be written as

$$e = [k_p C_{sp} + (1-k_p)C_v]T = C_{vm}T, \tag{7}$$

where $C_{sp}$ is the specific heat of the solid particles, $C_v$ is the specific heat of the gas at constant volume and $C_{vm}$ is the specific heat of the mixture at constant volume. For equilibrium conditions, the specific heat of the mixture at constant pressure is

$$C_{pm} = k_p C_{sp} + (1-k_p)C_p, \tag{8}$$

where $C_p$ is the specific heat of the gas at constant pressure. The ratio of the specific heats of the mixture is then given as

$$\Gamma = \frac{C_{pm}}{C_{vm}} = \frac{\gamma + \delta \beta_{sp}}{1 + \delta \beta_{sp}}, \tag{9}$$

where $\gamma = c_p/c_v$ is the specific heat ratio of the gas, $\beta_{sp} = C_{sp}/C_v$ is the specific heat ratio of solid particles and $\delta = k_p/(1-k_p)$. It is notable that $\Gamma$ decreases with increase in $k_p$, however, it increases with increase in values of $\gamma$.

Eliminating the temperature from Eqs. (4), (6) and (7), we may now write the specific internal energy of mixture as follows:

$$e = \left(\frac{1-Z}{\Gamma - 1}\right)\frac{p}{\rho}, \tag{10}$$

For isentropic change of state of the gas-solid particle mixture and thermodynamic equilibrium condition, we can calculate the so-called equilibrium speed of sound of the mixture for a given $k_p$ by using the effective ratio of specific heats and effective gas constant $R_M = (1-k_p)R_i$ as follows:

...

$$a = \left(\frac{dp}{d\rho}\right)_S^{1/2} = \left(\frac{\Gamma}{(1-Z)}\frac{p}{\rho}\right)^{1/2} = \left[\frac{\Gamma(1-k_p)R_iT}{(1-Z)^2}\right]^{1/2}, \quad (11)$$

where subscript '$s$' refers to the process of constant entropy. The initial speed of sound $a_o$ of the mixture is defined as

$$a_o^2 = \frac{\Gamma p_o}{(1-Z_o)\rho_o}. \quad (12)$$

The deviation of the behavior of a dusty gas from that of a perfect gas is indicated by the adiabatic compressibility of the mixture and is defined [23] as

$$\tau = \frac{1}{\rho}\left(\frac{\partial \rho}{\partial p}\right)_S = \frac{(1-Z)}{\Gamma p}, \quad (13)$$

where $(\partial \rho/\partial p)_S$ denotes the derivative of $\rho$ with respect to $p$ at the constant entropy.

The volume of solid particles lowers the compressibility of mixture, while the mass of solid particles increases the total mass, and therefore may add to the inertia of mixture. This can be shown in two limiting cases of mixture at the initial state. For $G = 1$, it follows from the Eqs. (6) and (4) that $Z_o = k_p$, $\rho_o = p_o/R_iT$ and $\tau = (1-k_p)/\Gamma p_o$, i.e. the presence of solid particles linearly lowers the compressibility of mixture in the initial state. In the other limiting case, i.e. for $G \to \infty$, the volume of the solid particles $V_{sp}$ tends to zero. According to Eq. (6), the volume fraction $Z_o$ is equal to zero. In this case, the compressibility $\tau = 1/\Gamma p_o$ is not affected by the dust loading. The solid particles contribute only to increasing the mass and inertia of mixture. Further, the expression for the change-in-entropy across the shock front is given by Anand [20] as follows:

$$\Delta s = C_{vm}\ln(p/p_o) - C_{pm}\ln(\rho/\rho_o) + C_{pm}\ln[(1-Z)/(1-Z_o)], \quad (14)$$

where $C_{vm} = (1-k_p)R_i/(\Gamma-1)$, and $C_{pm} = \Gamma(1-k_p)R_i/(\Gamma-1)$.

Using Eq. (10), Eq. (3) transforms into

$$\frac{\partial p}{\partial t} + u\frac{\partial p}{\partial r} + \rho a^2\left(\frac{\partial u}{\partial r} + D\frac{u}{r}\right) = 0. \quad (15)$$

Now, let us consider a shock wave propagating into a homogeneous mixture of a perfect gas and spherically small solid particles. In a frame of reference moving with the shock

___

front, the jump conditions at the shock are given by the principles of conservation of mass, momentum and energy across the shock, namely,

$$\rho(U - u) = \rho_o U, \tag{16}$$

$$p + \rho(U - u)^2 = p_o + \rho_o U^2, \tag{17}$$

$$e + \frac{p}{\rho} + \frac{(U-u)^2}{2} = e_o + \frac{p_o}{\rho_o} + \frac{U^2}{2}, \tag{18}$$

where $U$ and $u$ are, respectively, the shock front propagation velocity and the velocity of mixture. The flow quantities with the suffix '$o$' and without suffix denote, respectively, the values of flow quantities in upstream region, i.e. ahead of shock front and in downstream region, i.e. behind of shock front. Also, effects due to viscosity and thermal conductivity are omitted and it is assumed that the dusty gas has an infinite electrical conductivity. The upstream Mach number $M$, which characterizes strength of shock, is defined as

$$M = U/a_o, \tag{19}$$

The shock jump relations for pressure, temperature, density and velocity of mixture in terms of $M$ are given by Anand [20] as

$$\frac{p}{p_o} = \frac{2\Gamma M^2 - (\Gamma - 1)}{(\Gamma + 1)}, \tag{20}$$

$$\frac{T}{T_o} = \frac{[2M^2\Gamma - (\Gamma - 1)][2(1 - Z_o) + M^2(\Gamma - 1 + 2Z_o)]^2}{(\Gamma + 1)^2[M^2(\Gamma - 1) + 2]M^2}, \tag{21}$$

$$\frac{\rho}{\rho_o} = \frac{(\Gamma + 1)M^2}{M^2(\Gamma - 1 + 2Z_o) + 2(1 - Z_o)}, \tag{22}$$

$$\frac{u}{a_o} = \frac{2(1 - Z_o)(M^2 - 1)}{(\Gamma + 1)M}. \tag{23}$$

## 3. The geometrical shock dynamics theory and analytical solutions

In this section, we developed the geometrical shock dynamics model to provide a simplified, approximate treatment for the propagation of imploding shock waves in a two-phase mixture of a perfect gas and small solid particles. The geometrical shock dynamics [22] provides practically accurate results especially for continuously accelerating imploding shock waves. Consequently, the present model is well suited for studying shocks in the self-propagating limit, in which the front is moving steadily or accelerating.

According to the geometrical shock dynamics approach, the characteristic form of the governing Eqs. (1), (2) and (15), is easily obtained by forming a linear combination of Eqs. (1) and (15) in only one direction in $(r,t)$-plane. The linear combination of these two equations can be written as

$$\frac{\partial p}{\partial t}+(u+\lambda)\frac{\partial p}{\partial r}\pm\lambda\rho\frac{\partial u}{\partial t}+\rho(a^2+u\lambda)\frac{\partial u}{\partial r}+D\rho\, a^2\frac{u}{r}=0, \qquad (24)$$

The conditions that this combination involves the derivatives in only one direction, are given by

$$\frac{\partial p}{\partial t}=(u+\lambda)\frac{\partial p}{\partial r} \text{ or } \frac{\partial r}{\partial t}=(u+\lambda), \qquad (25)$$

and $\lambda\rho\frac{\partial u}{\partial t}=\rho(a^2+\lambda u)\frac{\partial u}{\partial r}$ or $\lambda\frac{\partial r}{\partial t}=a^2+\lambda u$, (26)

Eqs. (25) and (26) give

$$\lambda=\pm a \text{ i.e., } \frac{\partial r}{\partial t}=u\pm a, \qquad (27)$$

It shows the fact that the characteristic curves in $(r,t)$-plane represent the motion of possible disturbances whose velocity differs from the velocity of mixture $u$ by the value $\pm a$ (speed of sound), respectively, for diverging and converging shock waves. Now, Eq. (24) can be written as

$$\frac{\partial p}{\partial t}+(u\pm a)\frac{\partial p}{\partial r}\pm a\rho\frac{\partial u}{\partial t}\pm\rho a\,(u\pm a)\frac{\partial u}{\partial r}+D\rho\, a^2\frac{u}{r}=0, \qquad (28)$$

The above Eq. (28) is exact and holds throughout the flow since it is just a combination of the basic Eqs. (1), (2) and (15). By using above Eq. (28) we may write the characteristic form of the governing Eqs. (1), (2) and (15), i.e. the form in which equation contains derivatives in only one direction in the $(r,t)$-plane, as

$$dp+\rho\, a\, du+D\rho\, a^2\frac{u}{u+a}\frac{dr}{r}=0 \quad \text{along } C_+ \text{ i.e. } \frac{dr}{dt}=u+a, \qquad (29)$$

and

$$dp - \rho\, a\, du + D\rho\, a^2 \frac{u}{u-a}\frac{dr}{r} = 0 \quad \text{along} \quad C_- \quad \text{i.e.} \quad \frac{dr}{dt} = u - a, \tag{30}$$

Eqs. (29) and (30) represent the characteristic equations for exploding and imploding shock waves, respectively. The geometrical shock dynamics approach states that when relevant equations are written first in the characteristics form, the differential relation which must be satisfied along a characteristic can be applied to the flow quantities just behind the shock front. Together with the shock jump relations, this rule determines the propagation of the shock waves. We assume here that the shock jump relations to hold, of course, within the order of approximation determine by a constant value of $U$. We apply here the differential relation (30) along the negative characteristic $C_-$ behind the shock wave. Together with the shock jump relations, we are able to describe the shock velocity or the related quantities in terms of the quantities just ahead of the shock front. Eq. (30) is valid only along the negative characteristic curve $C_-$ in the $(r, t)$-plane, behind the imploding shock front.

The idea of the characteristic rule of Whitham [22] is to apply, on the negative characteristic curve $C_-$ along the imploding shock front. We thus neglect the difference in the constants of integration obtained when Eq. (30) is solved on different characteristics that intersect the shock front. These differences arise from the non-uniformity of the flow behind the shock, so the characteristic rule effectively ignores the influence of the flow behind the shock wave on the shock propagation. Because the effect of the flow behind the shock on the shock dynamics is ignored, the method is very good for situations where the shock wave accelerates with time, so that features of the flow behind do not 'catch up' with the shock. The excellent examples of flows with this characteristic are converging shock waves, which are the subject of this paper.

Now, assuming that the negative characteristic curve $C_-$ applies on the shock front, we can use the shock jump relations given by Eqs. (20) – (23) to write the quantities in it, which are those immediately behind the shock front, in terms of those ahead of shock front and Mach number. The shock jump relations we use here are the shock conditions for the dusty gas [20] rather than the shock conditions for an ideal gas used by Whitham [21].

___________________________________________________________________________

Now, substituting the shock jump relations given by Eqs. (20) – (23) into the negative characteristic curve $C_-$ we get a first order ordinary differential equation in $M$ as

$$\lambda(M)\frac{dM}{M} + D\frac{dr}{r} = 0, \tag{31}$$

where $\lambda(M) = \dfrac{\phi_1(M)}{\phi_2(M)\,\phi_3(M)}$,

$\phi_1(M) = (M^2 - 1)\{2\Gamma M^2 - (\Gamma - 1)\}\{M^2(\Gamma - 1 + 2Z_o) + 2(1 - Z_o)\}$,

$\phi_2(M) = 2(1 - Z_o)(M^2 + 1)[\{(\Gamma - 1)M^2 + 2\} - \{M^2(\Gamma - 1 + 2Z_o) + 2(1 - Z_o)\}\psi]$,

$\phi_3(M) = 2M^2 - (M^2 + 1)\psi$, and

$\psi(M) = \sqrt{[2\Gamma M^2 - (\Gamma - 1)]/[(\Gamma - 1)M^2 + 2]}$.

On integration, Eq. (31) discloses a relation between the propagation distance and Mach number as

$$r(M)K' = \exp\left[-\frac{1}{D}\int \lambda(M)\frac{dM}{M}\right], \tag{32}$$

where $K'$ is a constant of integration.

The variation of the function $\lambda(M)$ with the upstream Mach number $M$ for $\beta_{sp} = 1$, $\gamma = 7/5$ and various values of $k_p$ and $G$ are shown in Fig. 2. It is obvious that $\lambda(M)$ varies little for small values of $M$ and remains unaffected for large values of $M$. Thus, the propagation distance-Mach number relation for a dusty (32) may be easily written as
$M = Kr^{-D/\lambda(M)}$, where $K$ is a constant.

Thus, the analytical expression for the non-dimensional propagation velocity of shock may be written as

$$\frac{U}{a_o} = Kr^{-D/\lambda(M)}. \tag{33}$$

This equation is valid for the imploding shock waves in the mixture of a perfect gas and small solid particles and is the main result of the present study. Thus, the characteristic rule for the propagation velocity is $U \propto r^{-1/\lambda(M)}$ for cylindrical shocks and is $U \propto r^{-2/\lambda(M)}$ for spherical shocks, where $r$ is the radius of shock front. It is worth

___

mentioning that for strong diverging shock waves in a perfect gas, Whitham [22 page273] obtained the characteristic rule as $U \propto r^{-1/n}$ for cylindrical shock waves and $U \propto r^{-2/n}$ for spherical shock waves, where $n = (1 + 2(1-\mu^2)/(\gamma+1)\mu)(1+2\mu+1/M^2)$, and $\mu^2 = ((\gamma-1)M^2 + 2)/(2\gamma M^2 - (\gamma-1))$. The characteristic rule may be used for investigating the nature and behavior of the flow variables behind the imploding shock waves in the two-phase gas-particle flows.

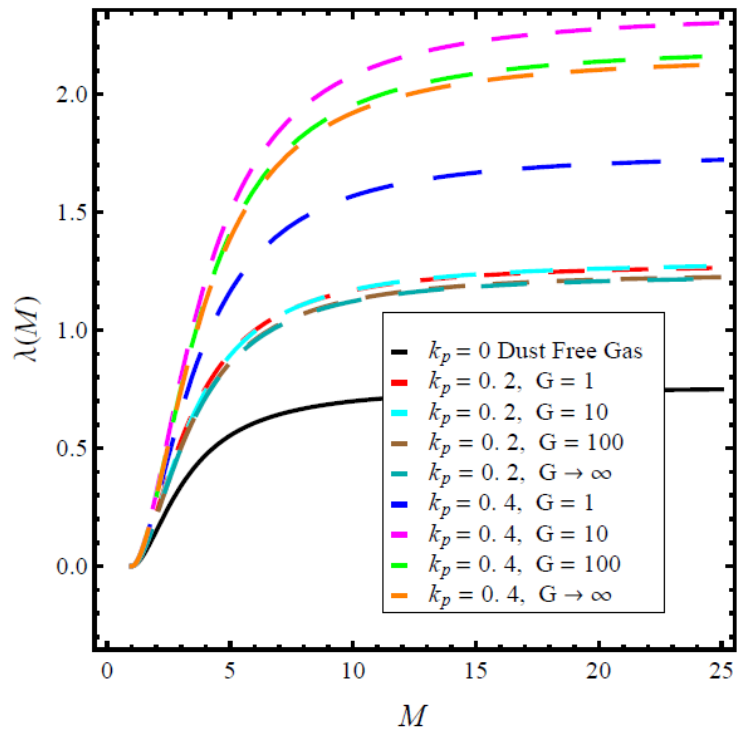

**Fig.2** The variation of $\lambda(M)$ with $M$ for various values of $k_p$ and $G$.

Now, the corresponding analytical expressions for the distribution of pressure $p$, temperature $T$, density $\rho$, velocity of mixture $u$, and speed of sound $a$ just behind the imploding shock front can be easily written as

$$\frac{p}{p_o} = \frac{2K^2 \Gamma r^{-2D/\lambda} - (\Gamma-1)}{(\Gamma+1)}, \qquad (34)$$

$$\frac{T}{T_o} = \frac{[2K^2\Gamma r^{-2D/\lambda} - (\Gamma-1)][2(1-Z_o) + K^2(\Gamma-1+2Z_o)r^{-2D/\lambda}]^2 r^{2D/\lambda}}{K^2(\Gamma+1)^2[(\Gamma-1)K^2 r^{-2D/\lambda} + 2]}, \qquad (35)$$

$$\frac{\rho}{\rho_o} = \frac{(\Gamma+1)}{(\Gamma-1+2Z_o)+2K^2(1-Z_o)r^{2D/\lambda}}, \tag{36}$$

$$\frac{u}{a_o} = \frac{2(1-Z_o)(K^2 r^{-2D/\lambda}-1)r^{D/\lambda}}{K(\Gamma+1)}, \tag{37}$$

$$\frac{a}{a_o} = \frac{2(1-Z_o)(K^2 r^{-2D/\lambda}-1)r^{D/\lambda}}{K(\Gamma+1)}. \tag{38}$$

Further, the analytical expression for the adiabatic compressibility of mixture just behind the imploding shock front is obtained by using Eqs. (13) and (34) as

$$\tau(p_o) = \frac{(1-Z)(\Gamma+1)}{\Gamma\left(2K^2\Gamma r^{-2D/\lambda}-(\Gamma-1)\right)}. \tag{39}$$

Finally, the analytical expression for the change-in-entropy across the imploding shock front in a dusty gas flow is easily obtained by using Eqs. (14), (34) and (36) as:

$$\begin{aligned}\frac{\Delta s}{R_i} &= \frac{(1-k_p)}{(\Gamma-1)}\ln\left[\frac{2K^2\Gamma r^{-2D/\lambda}-(\Gamma-1)}{(\Gamma+1)}\right]\\ &-\frac{\Gamma(1-k_p)}{(\Gamma-1)}\ln\left[\frac{(\Gamma+1)}{(\Gamma-1+2Z_o)+2K^2(1-Z_o)r^{2D/\lambda}}\right]\\ &+\frac{\Gamma(1-k_p)}{(\Gamma-1)}\ln\left[\frac{1-Z}{1-Z_o}\right]\end{aligned} \tag{40}$$

Thus, the influence of the mass concentration of solid particles and dust loading parameter on the shock velocity, the flow field quantities and the change-in-entropy can be explored from the above analytical expressions (33) – (40).

## 4. Results and discussion

In the present paper the general analytical solution for imploding shock waves in a two-phase mixture of a perfect gas and small solid particles was obtained by adopting the geometrical shock dynamics, due to Whitham [22] and further the general solution was examined and explored for cylindrical and spherical shock waves. The goal of the present investigation was to examine the influence due to dust particles on the imploding shock waves as they focus at the centre of convergence and the region of flow field immediately behind the imploding shock front. The general characteristic rule for the propagation velocity $U/a_o$ of imploding shock waves in a dusty gas is given by Eq. (33). The non-

___________________________________________________________________________

dimensional analytical expressions for the distribution of pressure $p/p_o$, temperature $T/T_o$, density $\rho/\rho_o$, velocity of mixture $u/a_o$, speed of sound $a/a_o$, and adiabatic compressibility $\tau(p_o)$ of mixture immediately behind the imploding shock front are given by Eqs. (34) – (39), respectively. Finally, the analytical expression for the change-in-entropy $\Delta s/R_i$ across the imploding shock front in the dusty gas is given by Eq. (40). These analytical expressions were derived by assuming that the disturbances due to the reflections, wave interactions in the wake, etc. do not overtake the imploding shock waves. It is worth mentioning that the analytical expressions for the shock velocity, pressure, temperature, density, velocity of mixture, speed of sound, adiabatic compressibility of medium and change-in-entropy across the shock front are functions of the propagation distance $r$ from the origin O, i.e. the centre of convergence, the upstream Mach number $M$, the mass fraction (concentration) of solid particles $k_p$ in the mixture, the dust loading parameter, i.e. the ratio of the density of the solid particles to the initial density of the gas $G$, the specific heat ratio of the solid particles $\beta_{sp}$ and the specific heat ratio of the gas $\gamma$. It is noteworthy that the effects due to the small solid particles enter through the parameters such as the mass fraction of solid particles in the mixture, the ratio of the density of the solid particles to the initial density of the gas and the specific heat ratio of the solid particles. It is notable that the analytical solutions for the imploding cylindrical and spherical shock waves in a dusty gas are principally identical and only a dimensionality index $D$ differs the two cases. The numerical computations of the flow variables behind the imploding shock front are carried out using MATHEMATICA and MATLAB codes. For the purpose of numerical calculations, the typical values of the specific heat ratio of the solid particles and the ratio of specific heats of the gas are taken to be 1 and 7/5, respectively. The value of $\beta_{sp}=1$ and $\gamma=7/5$ corresponds to the mixture of air and glass particles [15]. In our analysis, we have assumed the initial volume fraction of solid particles $Z_o$ to be a small constant. From Fig. 1 it is obvious that the values of $k_p$ from 0 to 0.4 with $G$ from 1 to $\infty$ give small values of $Z_o$, in general. Therefore, the values of the constant parameters are taken to be $\gamma=7/5$, $\beta_{sp}=1$, $k_p=0$, 0.2, 0.4, $G=1, 10, 100, \infty$, and $M=5$ for the general purpose of numerical computations.

___

The value of $k_p = 0$ corresponds to the case of a perfect gas or dust-free gas. The value $G = 1$ corresponds to $Z_o = k_p$, i.e. the case when initial volume fraction of solid particles in the mixture is equal to the mass fraction of solid particles. It is very useful to mention that the present analysis serves an analytical description for the propagation of imploding shock waves through an in-viscid, non-heat conducting and electrically infinitely conducting mixture of a perfect gas and small solid particles. In the present investigation, the two cases are considered-cylindrical and spherical converging shocks to highlight the differences between the 2D and 3D convergence.

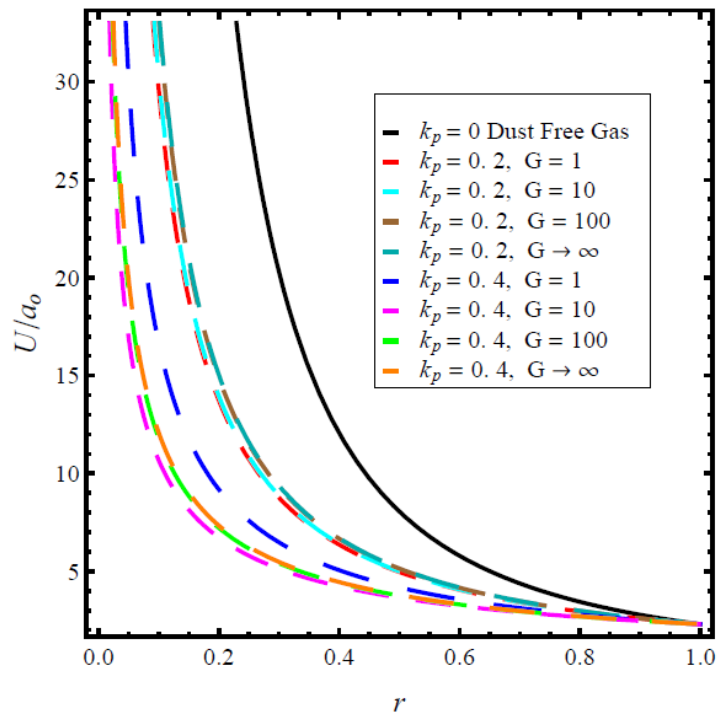

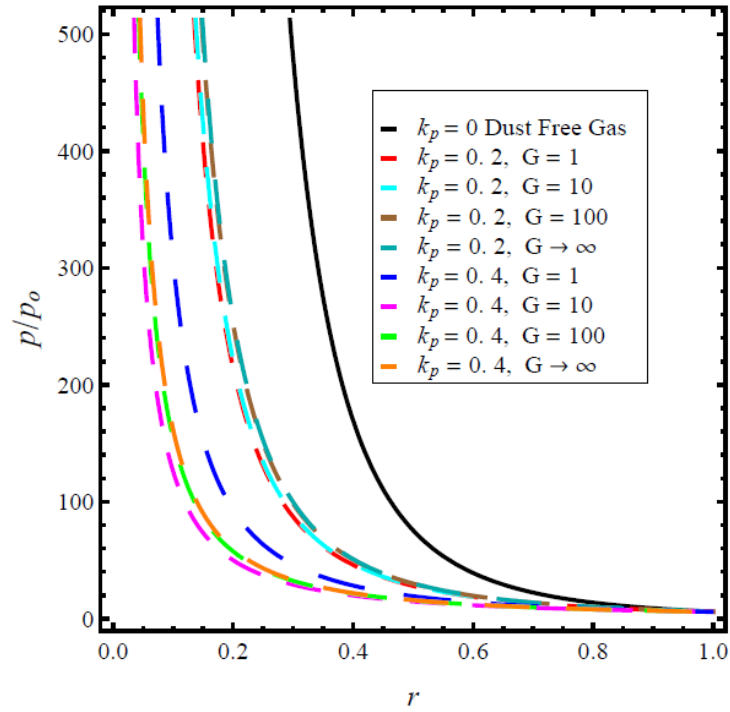

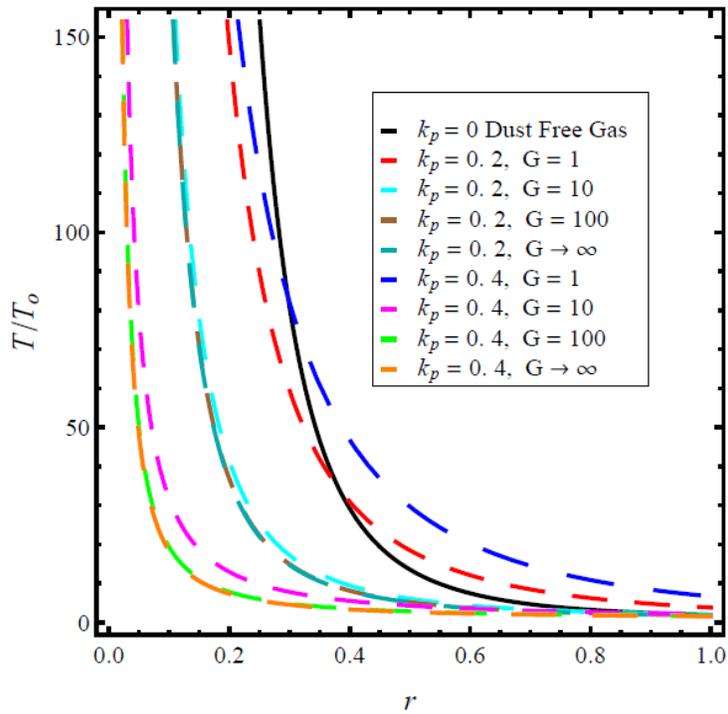

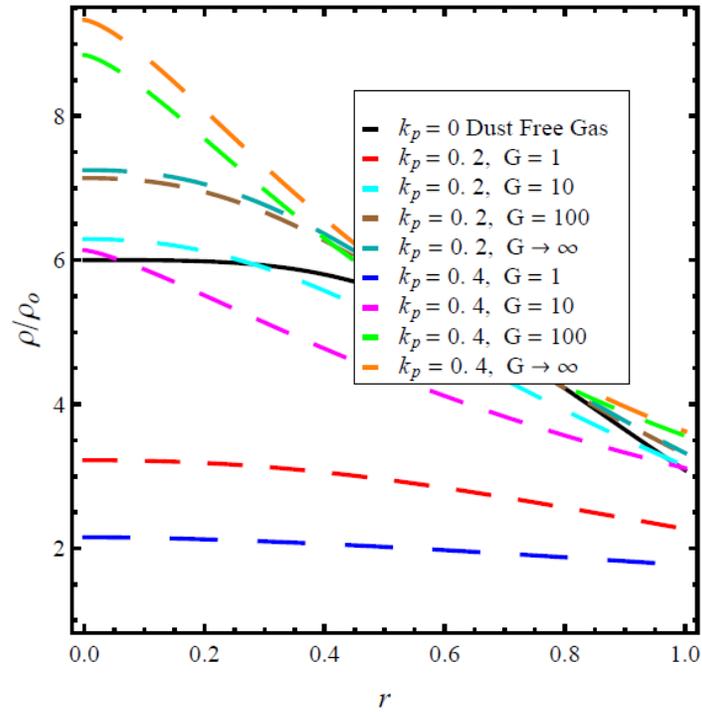

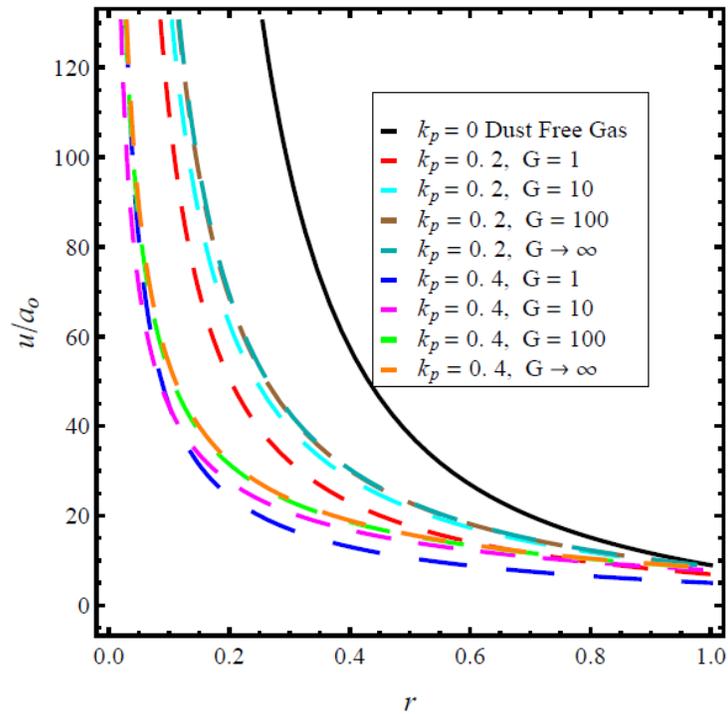

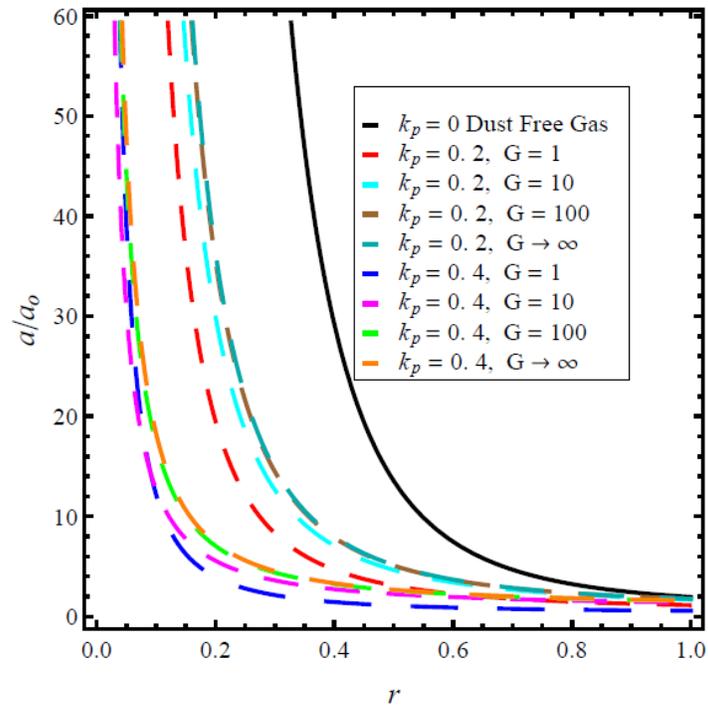

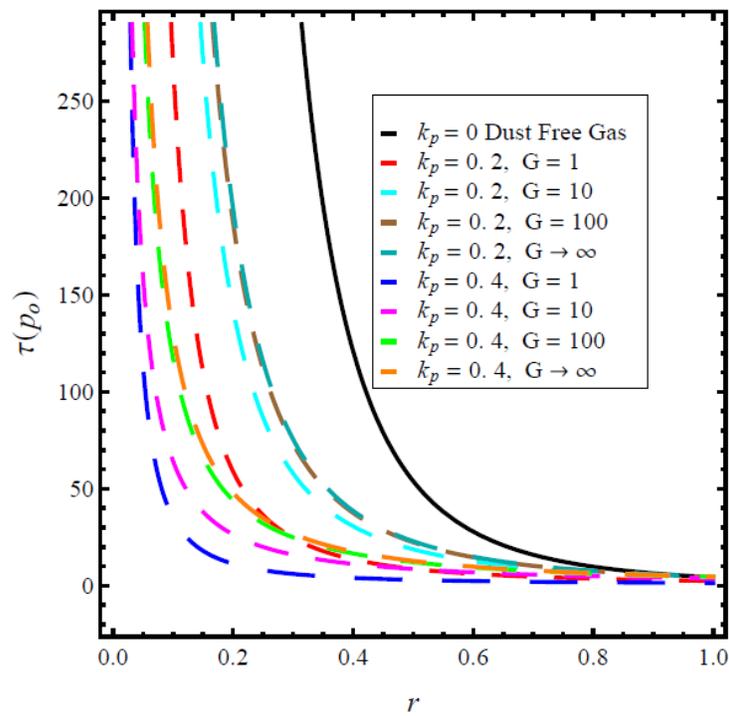

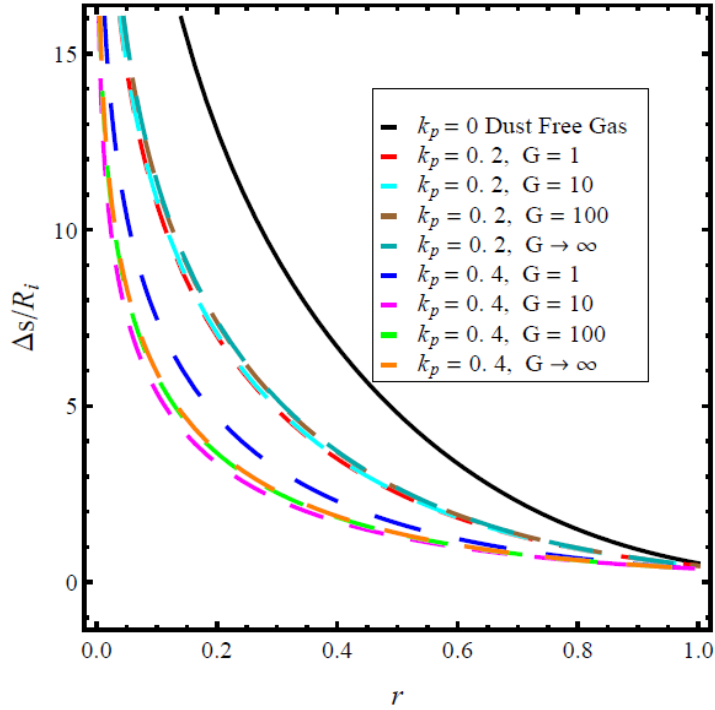

**Fig. 3** The variations of $U/a_o$, $p/p_o$, $T/T_o$, $\rho/\rho_o$, $u/a_o$, $a/a_o$, $\tau(p_o)$ and $\Delta s/R_i$ distribution behind the imploding cylindrical shock wave with $r$ for various values of $k_p$ and $G$.

The non-dimensional analytical expressions for the shock velocity, pressure, temperature, density, velocity of mixture, speed of sound, adiabatic compressibility and change-in-entropy behind the imploding cylindrical and spherical shock front are, respectively, obtained by taking the dimensionality index $D = 1$ and $D = 2$ in the Eqs. (33) – (40). In numerical computations, the value of constant $K = 2.30014659$ for cylindrical shock waves and $K = 1.05813487$ for spherical shock waves, which is obtained by assuming that $U = 5a_o$ at $r = 0.5$ for $\beta_{sp} = 1$, $\gamma = 7/5$, $M = 5$, $k_p = 0.2$ and $G = 10$. The variations of the shock velocity $U/a_o$, pressure $p/p_o$, temperature $T/T_o$, density $\rho/\rho_o$, velocity of mixture $u/a_o$, speed of sound $a/a_o$, adiabatic compressibility of mixture $\tau(p_o)$ and change-in-entropy distribution $\Delta s/R_i$ behind the imploding cylindrical shock front and spherical shock front with the propagation distance $r$ for $\beta_{sp} = 1$, $\gamma = 7/5$, $M = 5$ and various values of $k_p$ and $G$ are, respectively, shown in Figs. 3 and 4. It is observed from Figs. 3 and 4 that the shock velocity, pressure, temperature, velocity of mixture, speed of sound, adiabatic compressibility and change-in-entropy increase rapidly and tend to infinity as the cylindrical or spherical shock wave approaches at the axis or centre of convergence. It is notable that the distribution of density immediately behind the cylindrical (or spherical) shock front increases as the shock wave moves towards the axis (or centre) of convergence except for $G=1$, especially with $k_p = 0.4$. It is notable that the velocity of cylindrical and spherical shock decrease with increase in the mass concentration of solid particles in the mixture. The velocity of cylindrical and spherical shocks increases with the dust loading parameter $G$ for $k_p = 0.2$, however, for $k_p = 0.4$ the velocity of cylindrical and spherical shocks first decreases up to $G=10$ and then it increases. This behavior of the velocity of cylindrical and spherical shocks, especially for the case of $k_p = 0.4, G = 10$ differs greatly from the case of a perfect (dust-free) gas. Similarly the pressure behind the cylindrical and spherical shock front decreases as the mass concentration of solid particles increases in the mixture. The pressure behind the cylindrical or spherical shock front increases with the dust loading parameter $G$ for $k_p = 0.2$. However, for $k_p = 0.4$, the pressure first decreases with increase in the value of dust loading parameter $G$ up to $G=10$ and then the pressure increases. This behavior of the pressure, especially for the case of $k_p = 0.4, G = 10$ differs greatly from the dust-free case. It is important to note that the temperature behind the cylindrical and spherical shock fronts decreases with the mass concentration of solid particles in the mixture. It is also notable that only for value of $G=1$, the temperature behind the cylindrical shock front increases with the mass concentration of solid particles in the mixture. An increase in the dust loading parameter $G$ leads to a decrease in the temperature behind the cylindrical and spherical shock fronts. This behavior of the temperature for both dimensionalities, especially for the case of $k_p = 0.4, G = \infty$ differs greatly from the dust-free case. The density behind the cylindrical or spherical shock front decreases with increase in the mass concentration of solid particles for the values of $G \leq 10$, however, the density increases for the values of $G \geq 100$. The density behind the cylindrical and spherical shock front increases with the

___________________________________________________________________________________

dust loading parameter $G$. This behavior of the density, especially for the case of $k_p = 0.4, G = 1$ differs greatly from the dust-free case. The velocity of mixture, the speed of sound and the adiabatic compressibility just behind the cylindrical or spherical shock front decrease with the mass concentration of solid particles in the two-phase flow of a perfect gas and small solid particles. It is notable that the velocity of mixture, the speed of sound and the adiabatic compressibility immediately behind the cylindrical shock front increase with increase in the value of dust loading parameter $G$. The velocity of mixture, speed of sound and adiabatic compressibility of mixture just behind the spherical shock front increase with the dust loading parameter $G$ for $k_p = 0.2$. However, for $k_p = 0.4$ the velocity of mixture, speed of sound and adiabatic compressibility first decrease with increase in the value of dust loading parameter $G$ up to $G = 10$ and then they increase. This behavior of the velocity of mixture, speed of sound and adiabatic compressibility, especially for the cases of $k_p = 0.4, G = 1$ (cylindrical shocks) and $k_p = 0.4, G = 10$ (spherical shocks) differs greatly from the dust-free case. The change-in-entropy across the cylindrical or spherical front decreases with increase in the mass concentration of solid particles in the mixture. It is obvious from Figs. 3 and 4 that an increase in the dust loading parameter $G$ leads to an increase in the change-in-entropy for $k_p = 0.2$. However, for $k_p = 0.4$ the change-in-entropy first decreases with increase in the value of dust loading parameter $G$ up to $G = 10$ and then it increases. This behavior of the change-in-entropy, especially for the case of $k_p = 0.4, G = 10$ differs greatly from the dust-free case.

It is notable that large changes are found in the values of flow variables for small values of the dust loading parameter $G$ and vice versa. The variations in the flow variables with the dust loading parameter $G$ are more pronounced at higher values of $k_p$. This may be physically interpreted as follows. At constant $k_p$, there is a substantial decrease in $Z_o$ with increase in $G$ (see Fig. 1), i.e. the volume fraction of solid particles in the undisturbed medium becomes, comparatively, very small. It is worth mentioning that in the case of $G = 1$, small solid particles with the density equal to that of the perfect gas in the mixture occupy a significant portion of the volume which decreases the

___________________________________________________________________________

compressibility of the medium remarkably. However, for example in the case of $G = 100$, small solid particles with the density equal to hundred times that of the perfect gas in the mixture occupy a very small portion of the volume and, therefore, compressibility is not reduced much; but, the inertia of the mixture is increased significantly due to the particle load. An increase in $k_p$ from 0.2 to 0.4 for $G = 100$ means that the perfect gas in the mixture constituting 80% of the total mass and occupying 99.75% of the total volume now constitutes 60% of the total mass and occupies 99.34% of the total volume. Due to this reason, the density of the perfect gas in the mixture is highly decreased, which overcomes the effect of incompressibility of the mixture and ultimately causes a decrease in the shock velocity, and the above-mentioned behavior of the flow variables. Figs. 3 and 4 illustrate similar effects for an increase in $k_p$ from 0.2 to 0.4 for other values of the dust loading parameter $G$. Thus, the volumetric fraction of the dust lowers the compressibility of mixture, however, the mass of the dust load may increase the total mass, and hence it may add to the inertia of mixture. Both effects due to the addition of dust, the decrease of mixture's compressibility and the increase of mixture's inertia may obviously influence the propagation of shock waves.

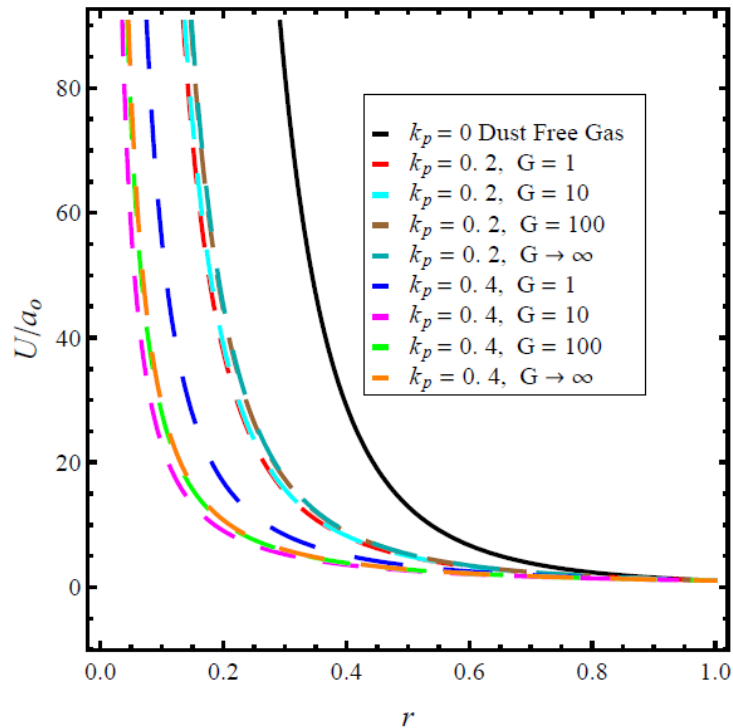

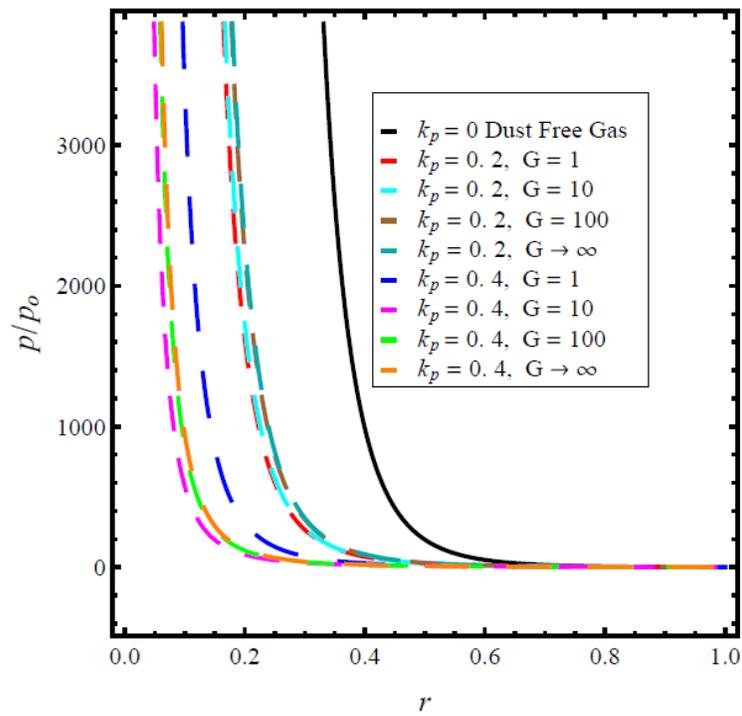

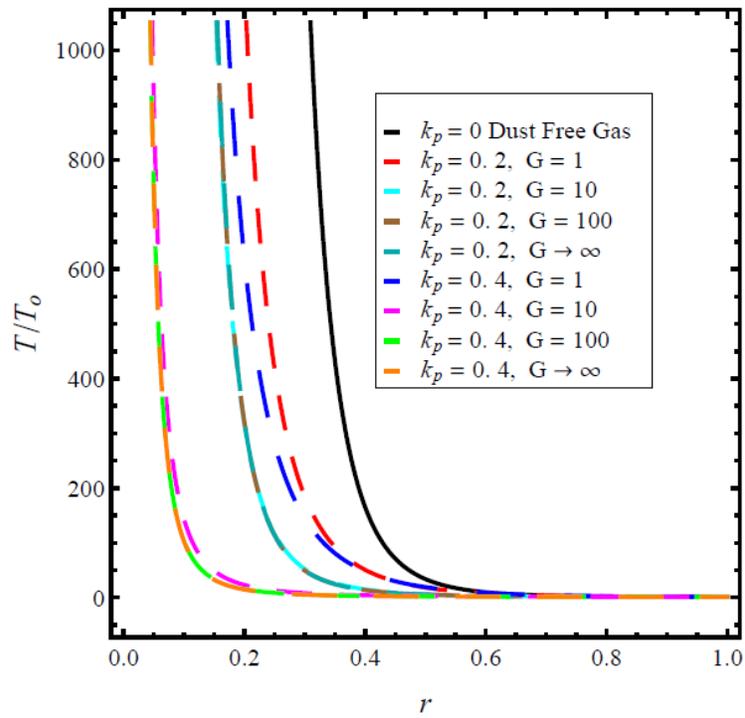

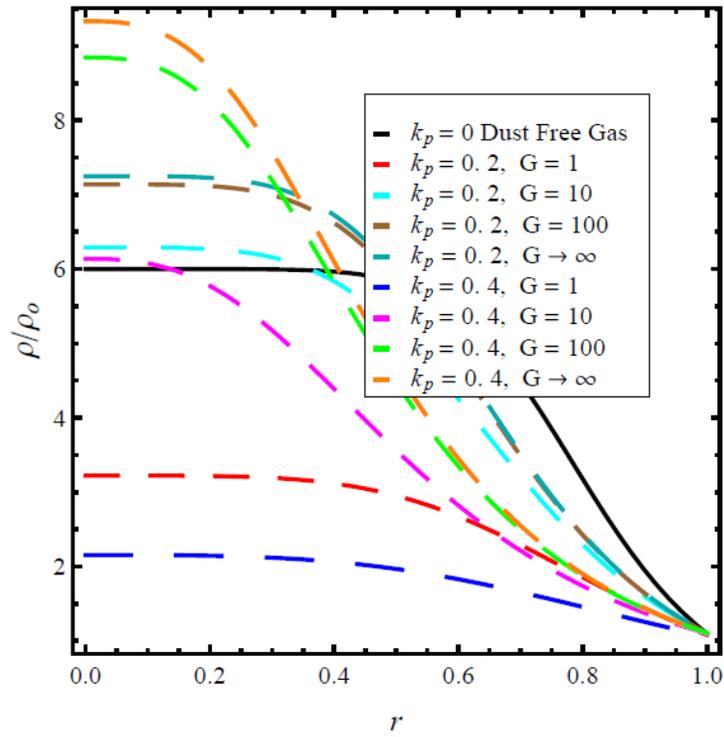

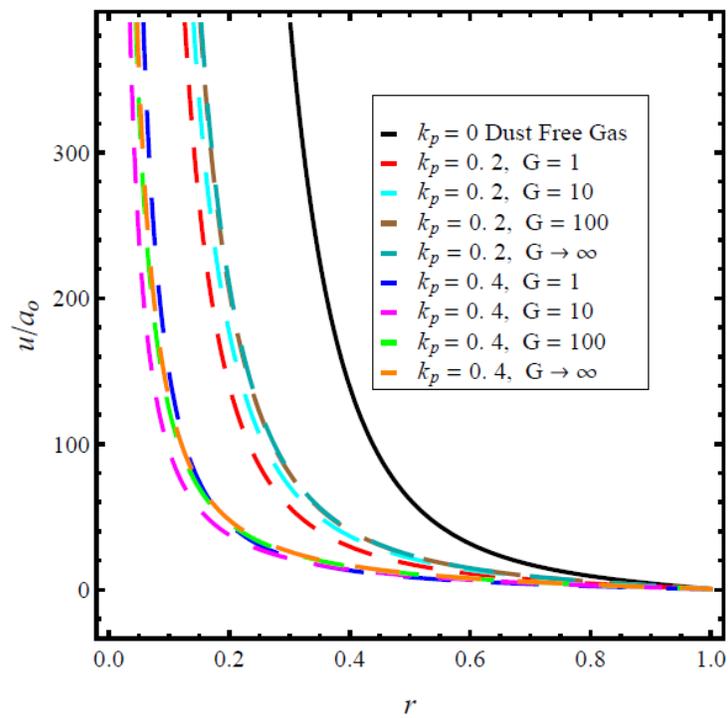

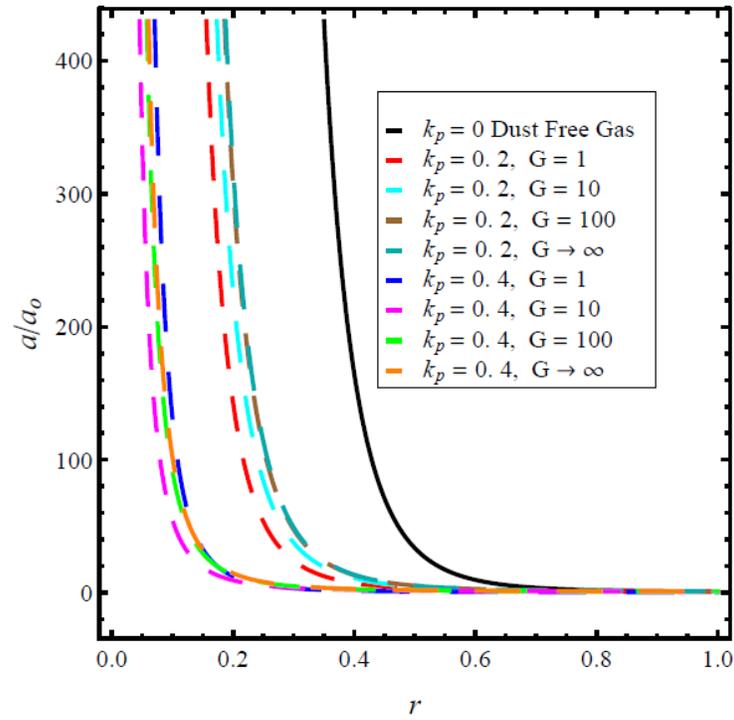

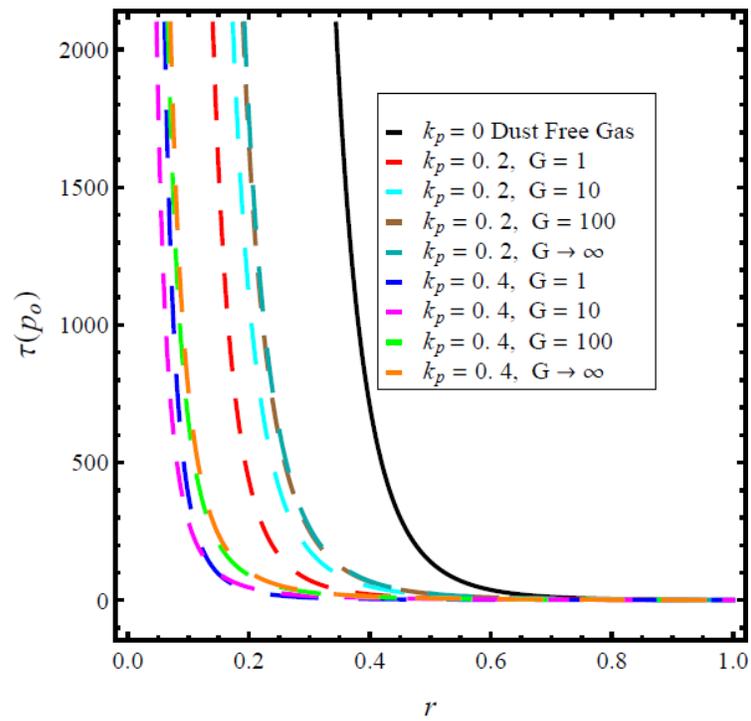

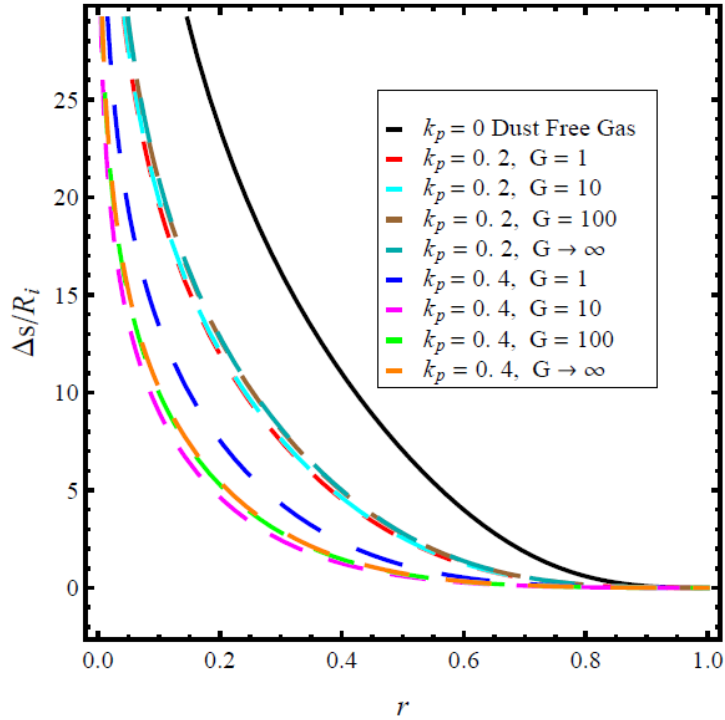

**Fig. 4** The variations of $U/a_o$, $p/p_o$, $T/T_o$, $\rho/\rho_o$, $u/a_o$, $a/a_o$, $\tau(p_o)$ and $\Delta s/R_i$ distribution behind the imploding spherical shock wave with $r$ for various values of $k_p$ and $G$.

It is worth mentioning that the effects of dust particles modify the numerical values of the shock velocity, pressure, temperature, density, velocity of mixture, speed of sound, adiabatic compressibility and change-in-entropy from their values for the perfect gas. However, the trends of variations of the shock velocity, pressure, temperature, density, velocity of mixture, speed of sound, adiabatic compressibility and change-in-entropy remain approximately unaffected, in general, for the cylindrical and spherical converging shock waves in a dusty gas. Thus, the present analysis provides a fairly accurate and complete description of the influence of dust particles on the shock velocity and flow variables behind the imploding cylindrical and spherical shock waves in a mixture of perfect gas and small solid particles.

## 5. Conclusions

The present work investigates the effects of dust loading parameters on the shock velocity and the flow field behind the imploding cylindrical and spherical shock waves propagating in a mixture of perfect gas and small solid particles. The following conclusions may be drawn from the findings of the current analysis:

1. The effects due to the mass concentration of solid particles and the dust loading parameter $G$, generally, do not change the trends of variations of the shock velocity and flow variables behind the cylindrical and spherical shock waves but they modify the numerical values of the shock velocity and flow variables from their values for the perfect (dust-free) gas case.
2. The shock velocity, pressure, temperature, velocity of mixture, speed of sound, adiabatic compressibility of mixture and change-in-entropy behind the cylindrical (or spherical) shock front increase as the shock wave focuses at the axis (or centre) of convergence. The density of mixture behind the cylindrical (or spherical) shock front increases as the shock wave focuses at the axis (or centre) of convergence except for $G = 1$.
3. The shock velocity, pressure, temperature, density, velocity of mixture, speed of sound, adiabatic compressibility and change-in-entropy behind the cylindrical and spherical shock fronts decrease with increase in the mass concentration of small solid particles in the mixture.
4. The shock velocity, pressure, density, velocity of mixture, speed of sound, adiabatic compressibility and change-in-entropy increase, however, the temperature decreases with an increase in the value of dust loading parameter $G$.
5. The trends of variations of the shock velocity and flow variables just behind the cylindrical or spherical shock waves in two-phase gas-particle mixture are similar to that of behind the cylindrical or spherical shock waves in a perfect gas.

The article concerns with the imploding problem, however, the methodology and analysis presented here may be used to describe many other physical systems involving non-linear hyperbolic partial differential equations. The potential applications of this study include simulation of problems in shock wave lithotripsy, explosive detonation and compressible flow in two-phase gas-particle fluids. This model may be used to describe

___________________________________________________________________________________

some of the overall features of the shock waves from underground nuclear explosion [24] and in the dusty environs of stars, stellar medium, double-detonation supernovae, etc. The model's advantages lie in the fact that it is capable of describing the flow field behind the imploding shock waves in an ideal gas as well as actual dusty environments. The model developed is of interest to the astrophysicists, fluid dynamicists, geophysicists and scientists working on imploding shock waves.